# Spin – orbit coupling and Lorentz force enhanced efficiency of $TiO_2$ based dye sensitized solar cells

*U. M. Kannan*, *M. Venkat Narayana*, *Ganesh Kotnana*, *Jaipal Kandhadi*, *L. Giribabu*, *Surya Prakash Singh* and *S. Narayana Jammalamadaka*[*]

U. M. Kannan, M. Venkat Narayana, Ganesh Kotnana, Dr. S. Narayana Jammalamadaka

Magnetic Materials and Device Physics Laboratory

Department of Physics

Indian Institute of Technology Hyderabad

Hyderabad – 502285

India

E-mail : surya@iith.ac.in

Jaipal Kandhadi , Dr. Surya Prakash Singh and Dr. L. Giribabu

Inorganic & Physical Chemistry Division

CSIR-Indian Institute of Chemical Technology

Hyderabad-500007

India

**Abstract**

We report on the effect of the strong spin – orbit coupling and the Lorentz force on the efficiency of $TiO_2$ based dye sensitized solar cells. Upon inclusion of $Ho_2O_3$, due to the strong spin – orbit coupling of the rare earth $Ho^{3+}$ ion, we do see 13% enhancement in the efficiency. We attribute such an enhancement in power conversion efficiency to the increased lifetime of the photo-excited excitons. Essentially, a $Ho^{3+}$ ion accelerates the phenomenon of the spin - rephasing or the intersystem crossing of the excitons in a photosensitizer. Increase in the absorbance and decrease in the photoluminescence intensity suggests a decrease in the recombination rate, hinting an enhanced charge transport and is in accordance with our electrochemical impedance spectra and the $J – V$ characteristics. From the above we strongly believe that enhanced efficiency of the device is due to increased intersystem crossing which would accelerate the exciton dissociation. On top of spin – orbit interaction, a configuration where the electric and magnetic fields are perpendicular to each other helped in enhancing the efficiency by 16%, suggesting that the Lorentz force also plays a dominant role in controlling the charge transport of the photo-generated charge carriers. We strongly believe that this simple and novel strategy of improving the efficiency may pave the way for realizing higher efficiency dye sensitized solar cells.

**Introduction**

Ever since the discovery of the dye – sensitized solar cells (DSSC) the demand for them has been rising due to their low production cost and easy fabrication. [1-5] However, the challenges involved in the commercial realization of the DSSC are manifold, ranging from the low efficiencies to the poor environmental stability of the devices.[6,7] Intense research has been carried out to improve the efficiencies by tailoring the organic photosensitizer,[8] photoelectrode modifications[9] and the interface engineering[10] for improving the device performance.

Though it has been suggested to tailor the photosensitizer and the morphology in order to improve the charge transport, it is found to be very difficult due to its complexity.[11] Another way of improving the charge transport is by increasing the effective life time of the photo-generated excitons and the diffusion lengths by having an intersystem crossing.[12] Essentially, in a molecule if an electron which exists in a singlet ground state (where electron spins are paired) is excited to a high energy state, there would be a probability of being either in an excited singlet state (paired with ground state) or in an excited triplet state. As the singlet ground state comprises a spin zero, the spin quantum selection rule precludes the excited triplet excitons to decay to the ground state, concurrently, triplet excitons cannot be excited by the photon absorption. In contrary, if there exists an excitonic transition between the singlet excited state (less life time) to the triplet excited state (more life time), diffusion length of the photo-generated excitons can be enhanced, which in turn improves the charge transfer mechanism between N719 dye LUMO to the $TiO_2$ acceptor states. Such process can be termed as the intersystem crossing/spin re-phasing. The relaxation for the spin selection rule through an intersystem crossing can take place by introducing the spin – orbit ($s – o$) coupling.[13] Over here the quantum state of the system is given by the total angular momentum, J = L + S, which may give weak spin forbidden bands with an overlap between the excited singlet excitonic levels and the excited triplet excitonic levels. Normally such an intersystem crossing enhances if the system consists of heavy atom molecules with a strong $s - o$ coupling and in the presence of paramagnetic elements.[14, 15]. Here, we make use of the Di-tetrabutylammonium cis-bis(isothiocyanato)bis(2,2'-bipyridyl-4,4'-dicarboxylato)ruthenium(II) photo sensitizer or commonly known as the N719 dye. The choice of N719 dye as the sensitizer for our studies has been motivated by the fact that the electron injection dynamics (EID) in the N719/$TiO_2$ system has been extensively studied in previous reports [13,16-18] and hence there is a general consensus on the bi-phasic injection kinetics from the N719 dye LUMO to the $TiO_2$ acceptor states. Essentially there exists an ultra-fast (≤ 100 fs) transfer component which involves a non-thermalized vibronic singlet state and a slower component which occurs in picosecond time-scales via thermalized triplet states. The internal relaxation via intersystem crossing to the triplet state occurs at a remarkably fast pace which is competing with the singlet injection component [19].

The Ru heavy atom in the N719 dye is known to introduce intersystem crossing between the excitonic levels with a very high yield. However, there are multiple arguments regarding the dominant injection channel between the excited dye state and the $TiO_2$ conduction band. Asbury *et. al,* [20] has observed that the ratio between short and long-lived states depend on the excitation wavelength (530 nm), where the main injection channel is from the non-thermalized singlet metal to ligand charge transfer state ($^1$MLCT state). However, Koops *et al* [13] concludes that triplet state injection is the dominant pathway for charge transport in high efficient devices. Hence, we intend to accelerate the intersystem crossing rate to improve the device performance.

As $Ho_2O_3$ (HO) is a powerful paramagnet [21] and the $Ho^{3+}$ consists of the strong $s – o$ interaction, [22] we believe that it may enhance the intersystem crossing and the charge transfer mechanism of the photo-generated excitons. For that reason, in the current work, we have incorporated a heavy rare earth oxide HO into the $TiO_2$ (TO) photo-anode to study the effect of the $s – o$ coupling on the performance of the TO based solar cell. In addition, HO is reported to have an excellent optical property like high absorption coefficient [23] which is also expected to support the light absorption in the photosensitized dye.

On top of that in recent past, the magnetic field effects have drawn tremendous interest to develop future energy conversion devices based on the semiconductors. [24, 25] Essentially, magnetic field would give an additional degree of freedom in addition to the electric field, which in turn gives an opportunity to amplify different heterogeneous electron transfer processes in addition to the charge transport, efficiency etc. Basically, if the charge carriers are experienced by a perpendicular electric and magnetic fields, they would be experienced by a Lorentz force which may further enhance the charge transfer mechanism. Keeping that in mind, we also made an attempt to see how the photo-generated excitons would behave by the simultaneous application of the external magnetic field as well as the $s - o$ interaction. We found that a moderate magnetic field of 100 Oe is producing a significant increase in the power conversion efficiency of the $TiO_2 + Ho_2O_3$ (HTO) device. As compared to the bare TO configuration, we could observe remarkable improvement in the charge transport for HTO based solar cells.

**Experimental**

Cross sectional view of the device is shown in **Figure. 2**. Initially, Fluorine doped Tin oxide (FTO) coated glass slides (Solaronix) were cleaned by an ultra-sonication method and using the DI water, acetone and isopropyl alcohol. 0.2 g of the TO paste (Solaronix) and 0.002 g of the HO (Sigma Aldrich) was thoroughly mixed and both the TO and the HTO (mixture of TO + HO) were deposited individually on separate FTO substrates using a blade coating technique. FTO substrates with the TO and the HTO were sintered in a furnace at 475°C for about 30 minutes. Both the sintered electrodes were then immersed in a solution of the Ru-bipyridyl dye (N719 dye) (Solaronix) and methanol ($3 \times 10^{-4}$ M) without any disturbance for about 24 hours in order for the dye loading in the TO/HTO network. On top of that the platinum (Pt) coated counter electrodes (Solaronix) were cleaned using an ultrasonication as stated above and sintered at 450°C for about 10 minutes. The dye soaked photoanodes were then cleaned using an ethanol and dried on a hotplate at 80°C for 5 minutes. Both the TO/HTO (photo anode) and the Pt (cathode) were combined together using binder clips and a drop of $I^-/I_3^-$ electrolyte solution (Solaronix) was added to the device. The final devices had an active area of 0.36 cm$^2$. The fabricated DSSC devices were subjected to the current density (*J*) vs. voltage (*V*) characterization under AM 1.5 conditions using an Oriel solar simulator, USA. The incident light power was standardized using a Si photodiode. The J(V) values were measured using a Keithley 2400 sourcemeter connected to a computer. The magnetization (*M*) vs. applied magnetic field ($\mu_0H$) measurement was carried out on the HTO film using a vibrating sample magnetometer. (7400 lakeshore VSM).

Morphological studies were carried out using a Scanning electron microscope (Carl Zeiss) using an accelerating voltage of 5kV. The optical absorbance was recorded using the thin film UV-Visible spectrophotometer (Shimadzu model UV -3600 spectrophotometer) and the photoluminescence spectroscopy (PL) of the dye coated photoanodes were recorded using Fluorolog-3 spectrofluorometer. The extinction coefficient (k) was measured using J A Woollam Co M2000 spectroscopic ellipsometer at three different angles 65°,70° and 75°. The electrochemical impedance spectrum (EIS) of the devices was recorded using an Iviumstat electrochemical interface under an applied bias voltage of 0.6 V and excitation signal of 10 mV amplitude over a frequency range from 100 mHz to 100 kHz.

**Results and discussion**

**Figure 1 (a)** shows the M vs. $\mu_0 H$ for the prepared thin film of HTO. It is evident from the graph that the M increases linearly with $\mu_0 H$ up to a maximum magnetic field of 20 kOe, hinting a paramagnetic in nature. Morphological studies were carried out using a SEM and Figure 1(b) shows the image pertinent to the morphology of the film at 34 KX magnification. It is apparent that the prepared film consists of a uniform surface with a mean roughness of about 6 μm, which was measured using an optical profilometer (Zeta). Cross sectional and the schematic of the experimental set up used for *J - V* characterization is shown in **Figure 2** (More details about the device fabrication are discussed in methods section)

Now we discuss about the characteristics of the TO and the HTO based solar cell devices with and without magnetic fields using a diode model for the DSSC devices. For that reason initially *J – V* characteristics of both the devices would be discussed without magnetic fields and the extracted results would be supported by the impedance spectroscopy, absorption spectra and photoluminescence spectra. Finally the characteristics of HTO based solar cell would be discussed by simultaneous application of both the spin – orbit coupling as well as the Lorentz force. Salient features of the present manuscript are (a) due to the strong *s - o* coupling there is an enhancement in the efficiency by 13% (b) absorbance increases while PL decreases for HTO based solar cells, which indicates the decrease in recombination rate (c) solar cell efficiency is further enhanced for low magnetic fields due to the presence of both the Lorentz force and the s – o coupling in the device.

The *J - V* characteristics for the fabricated device without magnetic field are shown in **Figure 3**. Left panel of Figure 3 shows the *J – V* characteristics and right panel shows the power density (P) vs. *V* characteristics. It is evident from the figure that the *J* consists maximum value at the zero voltage as it is in the short circuit condition. At higher potential, due to the exponential domination of the recombination resistance ($R_r$) over the series and the shunt resistances ($R_s$ & $R_{sh}$), *J – V* curves are modified from the maximum power voltage ($V_{MP}$) to the open circuit voltage ($V_{OC}$). On the other hand, the *P* shows zero at short circuit current density ($J_{SC}$) as *P = J * V* and peaks at the $V_{MP}$. Further increase in the *V* leads to decrease in the *P* due to the fact that the recombination current ($J_r$) increases which leads to decrease in the *J*. By assuming the cell voltage *V* as negative and the generated *J* as positive from the zero potential to $V_{OC}$, resultant *J* vs. *V* can be expressed by the general diode model equation for the DSSC device as given below:

$$J = J_L - J_0[\exp\{\frac{-q(V-JR_s)}{nk_BT}\} - 1] + \frac{(V-JR_s)}{R_{sh}} \qquad (1)$$

where $J_L$ is the light generated current density, $J_0$ is the reverse saturation current density, *n* is the diode ideality factor, $R_s$ is the series resistance, $R_{sh}$ is the shunt resistance, $k_B$ is the Boltzmann's constant, *q* is the fundamental charge and *T* is the temperature in Kelvin scale. Using the equation (1) and the *J –V* characteristics, $R_s$, $R_{sh}$, *n* and $J_0$ are calculated using an analytical parameter extraction procedure that has been proposed by Sarker *et. al*.[26] Calculated solar cell parameters and the corresponding equation that we used to extract are shown in **Table 1**.

It is apparent from table that the efficiency of the TO solar cell increases from 5.22 % to 5.90 % (~ 13% increase) in the presence of strong $s$ - $o$ interaction due to the presence of $Ho^{3+}$. We would like to correlate such an increase in efficiency through the increase in the rate of intersystem crossing from an excited singlet excitonic state to a triplet excitonic state of the N719 Dye in the presence of the strong $s – o$ coupling. **Figure 4** shows schematic of the photo-excitation of the spin states pertinent to the N719 Dye in the presence of the HO without magnetic field. Here, $S_0$ represents singlet ground state of the N719 dye. By impinging AM 1.5 spectrum on the HTO based solar cell, there would indeed be photo-generated excitons which consists of the electron hole pair in the singlet excited excitonic state ($S_1$) where they would have very less life time. External bias and band alignment of the N719 Dye with the TO leads to an exciton dissociation and an electron transfer from the singlet excited state to the conduction band ($CB$) of the TO respectively. However, the transition from the singlet excited state to the triplet excited state ($T_1$) is forbidden due to the quantum selection rule as we discussed before. In the present device as we have strong $s – o$ interaction due to the HO, there can be an excitonic transition from $S_1$ to $T_1$, which may give weak spin forbidden bands and hence the overlap between the singlet and the triplet excited states. Due to this process, there might be an enhancement in the triplet exciton population which leads to an increase in the exciton dissociation rate. Eventually this process may enhance the recombination time which in turn leads to an enhanced efficiency and is consistent with our results of the $J – V$ characteristics. On top of that as more excitons are dissociated in $T_1$ state, more electrons are driven to the external circuit through the conduction band of the TO and hence the increase in the short circuit current density as shown in **Table 1**.

The electrochemical impedance analysis (EIS) is performed to understand the charge transport variation that we encountered during the *J – V* characteristics of the HTO solar cell. **Figure. 5 (a)** shows the Nyquist plot and it is evident that the radius of the impedance arc decreases in the HTO based device, hinting the large charge transport within the medium, which is consistent with the variation of the *J* from the *J – V* characteristics. In addition, we have fitted the experimentally observed EIS data to a custom electrical circuit (inset Figure 5) which emulates the internal charge transport mechanism of the devices. From the fitted parameters, it is observed that there is a significant decrease in the charge transport resistance in the HTO system (~1.063 Ohm) with respect to TO (~ 4.96 Ohm). The same is realized using the EIS as a decrease in the radius of the impedance spectrum, hinting a perfect correlation with a decrease in the series resistance $R_s$ of HTO (7.69 ohm cm$^2$) with respect to TO (10.35 ohm cm$^2$) (see in Table 1). Earlier such variation of arc radius has been corroborated to the enhanced charge transport with in the medium.[27] Apart from the decrease in an arc radius from Figure 5, it is also evident that the TO based solar cell consists of two arcs, which suggests that the larger arc corresponds to the charge transport in the TO/Dye/I$^-$/I$_3^-$ and the smaller arc corresponds to the Warburg diffusion process of the I$^-$/I$_3^-$.[28] Figure 5(b) and Figure 5 (c) reveals the variation of the impedance and the phase angle with respect to the frequency. It is clear from both the graphs that the impedance and the phase angle decreases with respect to the frequency, suggesting that the real part dominates and capacitive part diminishes with respect to the frequency. This indicates a system with less capacitive effects like space charge effects and charge accumulation at the interfaces, which in turn helps the charge transport and the efficiency as observed through our *J – V* characteristics.

Striking coincidence between our *J – V* characteristics and EIS measurements motivated us to probe both the TO and the HTO devices further using the optical absorption and the photoluminescence studies. Essentially, we anticipate an increase in the absorption for the Dye after inclusion of the HO. Meeting our expectations, the optical absorption spectra (**Figure 6(a)**) from UV-Visible spectrophotometer shows an enhanced absorption at 527 nm which corresponds to the N719 Dye.[29] In addition to the above peak indeed there exists a peak at 344 nm which corresponds to the TO.[30] We attribute such an enhancement in the absorbance at 527 nm to the absorption of the visible light by the defect sites in the HO lattice which are also responsible for the colored appearance of the compound.[31]

In order to verify whether the increased absorbance is an inherent material property of HO or it is due to increase in dye loading, we calculated the absorption coefficient (α) as a function of the incident wavelength for both TO and HTO layers. The following equation was used to determine α

$$\alpha = \frac{4\Pi k}{\lambda} \quad (2)$$

where *k* is the extinction coefficient of the material and *λ* is the wavelength of incident radiation. Spectroscopic ellipsometry was performed on TO and HTO layers using J. A. Woollam Co M2000 spectroscopic ellipsometer. Measurements were performed at three different angles (65°, 70° and 75°) and the recorded psi and delta values were fitted with a B-spline optical model to generate refractive index (n) and extinction coefficient(k) as a function of wavelength (λ). The model generated *k* values were then used to calculate α according to equation (2). Inset of Figure 6(a) shows the variation of α with wavelength for both the TO and HTO devices. α (λ) curve for HTO lies much above the TO curve revealing the enhanced absorbance of the HTO device. A similar absorption enhancement was observed in the UV Visible absorbance spectrum in the wavelength range from 400 nm to 600 nm. As observed in the optical absorption spectra this improvement in the absorbance is solely due to the inclusion of HO in the DSSC device.

As we stated above, the $s$ - $o$ interaction enhanced excitonic dissociation can lead to a decrease in the intensity of the photoluminescence spectra. In order to realize such fact, we have carried out a systematic photoluminescence (PL) study (Fluorolog-3 spectrofluorometer) (Figure 6(b)) taking into account any possible emission from metal oxide or the solvent (methanol) as such. Essentially, PL data were recorded for four different configurations which include a solution of methanol (A), a solution of 0.002 g $Ho_2O_3$ (HO) in methanol (B) and blade coated films of HTO (C) and TO (D) on FTO coated glass slides. All the four samples were excited at 527 nm which is the absorption maximum for the N719 dye. Clearly we could observe an emission peak at 728.5 nm for samples C and D but not for the other samples. Similar emission wavelength was reported for N719 dye in earlier reports [32]. The red shift of the emission peak as compared to the dye absorption maximum is because the luminescence peak arises from the triplet excited state after intersystem crossing (ISC) from the vibronically excited singlet state. The quenching of luminescence intensity and decrease in full width at half maxima of HTO in comparison with TO is a clear evidence of the increase in intersystem crossing rate and an indication for the excitonic dissociation. Essentially, the $s$ – $o$ coupling induced intersystem crossing is effectively dissociated greater number of triplet excitons into charge carriers and the same is evident from the decrease in the PL emission intensity.

Simultaneous effect of the spin orbit interaction and the external magnetic field on the $J$ – $V$ characteristics of the HTO based solar cells is studied using a single pole electromagnet.

Schematic of the experimental setup is shown in **Figure 7**. The efficiency of the solar cell is measured in the presence of the three dimensional (3D) magnetic field, which means that the magnetic flux lines are both parallel and perpendicular to the plane of the device. The applied magnetic field was varied from 100 - 300 Oe in discrete steps of 100 Oe. The resultant *J-V* curves in the presence of different applied fields are shown in **Figure 8**. Inset shows the variation of the maximum power point with different magnetic field values. It is evident that the HTO device delivers the maximum operating power of all the fabricated devices at an applied magnetic field of 100 Oe. Indeed the device HTO at 100 Oe shows an efficiency of 6.06% which is an enhancement in the efficiency of 3% if we compare with the HTO without magnetic field and 16% enhancement with respect to the bare TO efficiency (5.22 %). Further increase in magnetic field leads to a decrease in efficiency, which suggests that low magnetic fields are beneficial to improve the efficiency. Essentially, an enhanced efficiency of about 3% at low magnetic fields for the HTO can be attributed to the predominant alterations of the local electronic surface properties of the HTO which can cause the slow electron recombination[19]. In addition to the intersystem crossing which leads to an excitonic dissociation, as there is a perpendicular magnetic field component with respect to the plane of the device, the charge carriers which are moving in the device may also be experienced by a forced called as the Lorentz force due to the presence of perpendicular electric and magnetic fields.[33] Mathematically it can be expressed as *F = q (E + v* x *B)*, where *F* is the Lorentz force, *q* is the electron charge, *E* is the electric field, *v* is the velocity of electron and *B* is the intensity of the magnetic field. From the Figure 7 it is very much clear that the electric field *E* is in (–*y*) – direction and the applied magnetic field *B* is in the *x* – direction. Due both E and B, the charge carriers would feel a force called Lorentz force, *v* x *B*. Due to this force, the electron cannot go to the circuit directly instead it follows a helical behavior as shown in the inset of Figure 7. As *v* is in *y* – direction and *B* is in *x* – direction and the electrons have negative charge, the force should be in *z* – direction. Due to the Lorentz force, the electrons would travel longer paths, which may enhance the recombination time and hence the efficiency. **Table 2** depicts the parameters that were extracted from the *J* – *V* characteristics in the presence of the weak magnetic field. It is evident that the efficiency increases for 100 Oe field and decreases for higher fields. The increase in efficiency for low magnetic fields can be attributed to the helical motion of the electrons due to the Lorentz force. In addition, due to the magnetic field the recombination time increases as a result of Lorentz force, which reflects as an increase in the fill factor (*FF*) value. The $R_s$ and $R_{sh}$ are characteristic values for a solar cell and for an ideal solar cell, $R_s$ should be less and $R_{sh}$ should be high in order to have higher efficiencies. From Table. 2, it is apparent that for the HTO at 100 Oe, the $R_s$ (2.46 ohm cm$^2$) is less and $R_{sh}$ (4.82 Kohm cm$^2$) is more, which suggests less resistance for the charge transport with in the device.

**Conclusion**

In the present study, we have successfully demonstrated the effect of a heavy rare earth oxide modified photoanode to improve the performance of dye sensitized solar cell based on TiO$_2$. The powerful paramagnetic HO could produce a 13 % enhancement in the power conversion efficiency of the DSSC. Such an enhancement in efficiency attributed to an increase in the intersystem crossing from the short lived singlet exciton to the long lived triplet excitons, which are accelerated due to the *s* - *o* coupling in the heavy Ho$^{3+}$ ion. Such an intriguing effect was supported by our PL measurements where we could observe a decrease in the emission intensity with HO addition. The improved charge transport within the device was affirmed by a low transport resistance as observed in the EIS measurements. Finally, the application of an external magnetic field as low as 100 Oe enhanced the device efficiency to about 6.06% due to the increase in recombination time as a result of the Lorentz force that was experienced by the charges. We believe that our work would indeed play a major role in developing DSSC devices based on magnetic phenomena.

**Acknowledgements**

We would like to thank Indian Institute of Technology Hyderabad for funding the research.

**References**

[1] B. O'regan, M. Grätzel, *Nature* **1991**, 353, 737.

[2] M. Grätzel, *J. Photochem. Photobiol. C* **2003**, 4.2 ,145.

[3] M. Law, L. E. Greene, J. C. Johnson, R. Saykally, P. Yang, *Nat. Mater*. **2005**, 4, 455.

[4] U. Bach, D. Lupo, P. Comte, J. E. Moser, F. Weissörtel, J. Salbeck, H. Spreitzer, and M. Grätzel, *Nature* **1998**, 395, 583.

[5] A. Shah, P. Torres, R. Tscharner, N. Wyrsch and H. Keppner, *Science* **1999**, 285, 692.

[6] A. Hagfeldt, G. Boschloo, L. Sun, L. Kloo, and H. Pettersson, *Chemical reviews* **2010,** 110, 6595.

[7] M. Ye, X. Wen, M. Wang, J. Iocozzia, N. Zhang, C. Lin and Z. Lin, *Materials Today* **2015,** 18, 155.

[8] S. Mathew, A. Yella, P. Gao, R. Humphry-Baker, B. F. E. Curchod, N. Ashari-Astani, I. Tavernelli, U. Rothlisberger, M. K. Nazeeruddin, and M. Grätzel, *Nature chemistry* **2014**, 6, 242.

[9] F. Sauvage, F. D. Fonzo, A. L. Bassi, C. S. Casari, V. Russo, G. Divitini, C. Ducati, C. E. Bottani, P. Comte, and M. Gratzel, *Nano letters* **2010**, 10, 2562.

[10] M. Graetzel, R. A. J. Janssen, D. B. Mitzi and E. H. Sargent, *Nature* **2012**, 488, 304.

[11] K. M. Lee, V. Suryanarayanan and K. C. Ho, *Sol. Energ. Mat. Sol. Cells* **2006**, 90, 2398.

[12] X. Sun, X. Wang, X. Li, J. Ge, Q. Zhang, J. Jiang and G. Zhang, *Macromol. Rapid Commun.* **2015**, 36, 298.

[13] S. E. Koops, B. C. O'Regan, P. R. F. Barnes and J. R. Durrant, *J. Am. Chem. Soc.* **2009**, 131, 4808.

[14] D. Beljonne, Z. Shuai, G. Pourtois and J. L. Bredas, *J. Phys. Chem. A* **2001**, 105, 3899.

[15] D. A. Skoog, F. J. Holler, and T. A. Nieman. *Principles of Instrumental Analysis*, 5th Ed. Brooks/Cole, 1998

[16] S. A. Haque, E. Palomares, B. M. Cho, A. N. M. Green, N. Hirata, D. R. Klug and J. R. Durrant, *J. Am. Chem. Soc.* **2005**, 127, 3456.

[17] B. Pandit, T. Luitel, D. R. Cummins, A. K. Thapa, T. Druffel, F. Zamborini and J. Liu, *J. Phys. Chem. A* **2013**, 117, 13513.

[18] M. Juozapavicius, M. Kaucikas, S. D. Dimitrov, P. R. F. Barnes, J. J. V. Thor and B. C. O'Regan, *J. Phys. Chem. C* **2013**, 117, 25317.

[19] M. Borgwardt, M. Wilke, T. Kampen, S. Mähl, W. Xiang, L. Spiccia, K. M. Lange, I. Y. Kiyan and E. F. Aziz, *J. Phys. Chem. C* **2015**, 119, 9099.

[20] J. B. Asbury, N. A. Anderson, E. Hao, X. Ai and T. Lian, *J. Phys. Chem. B* **2003**, 107, 7376.

[21] G. Adachi and N. Imanaka, *Chem. Rev* **1998**, 98, 1479.

[22] B. L. Rhodes, S. Legvold and F. H. Spedding, *Phys. Rev* **1958**, 109, 1547.

[23] T. Wiktorczyk, *Thin Solid Films* **2002**, 405, 238.

[24] F. Cai, S. Zhang, S. Zhou and Z. Yuan, *Chem. Phys. Lett.* **2014**, 591, 166.

[25] F. Cai, S. Zhang and Z. Yuan, *RSC Adv.* **2015**, 5, 42869.

[26] S. Sarker, H. W. Seo, K. S. Lee, Y. K. Jin, H. Ju and D. M. Kim, *Sol. Energy* **2015**, 115, 390.

[27] S. Sarker, H. W. Seo and D. M. Kim, *Chem. Phys. Lett.* **2013**, 585, 193.

[28] C. P. Hsu, K. M. Lee, J. T. W. Huang, C. Y. Lin, C. H. Lee, L. P. Wang, S. Y. Tsai and K. C. Ho, *Electrochim. Acta* **2008**, 53, 7514.

[29] M. K. Nazeeruddin, A. Kay, I. Rodicio, R. H. Baker, E. Müller, P. Liska, N. Vlachopoulos and M. Grätzel, *J. Am. Chem. Soc* **1993**, 115, 6382.

[30] H. Tang, K. Prasad, R. Sanjines, P. E. Schmid and F. Levy, *J. Appl. Phys* **1994**, 75, 2042.

[31] Y. Su, G. Li, X. Chen, J. Liu and L. Li, *Chem. Phys. Lett.* **2008**, 37, 762.

[32] R. Katoh and A. Furube, *J. Photochem. Photobiol. C* **2014**, 20, 1.

[33] D. J. Griffiths, *Introduction to electrodynamics*, 3rd Ed, p. no. 204, Prentice – Hall of India Pvt. Ltd, India.

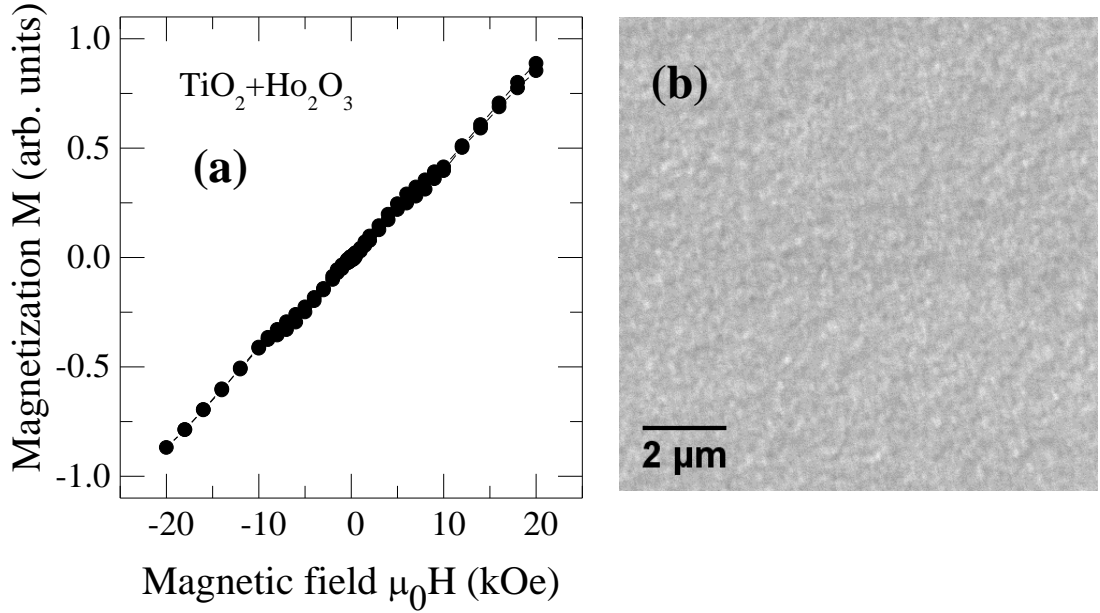

**Figure 1**: (a) Magnetic field vs. Magnetization graph for HTO, indicates a paramagnetic nature of the film. (b) SEM image of HTO. Roughness of the thin film is about 6 μm.

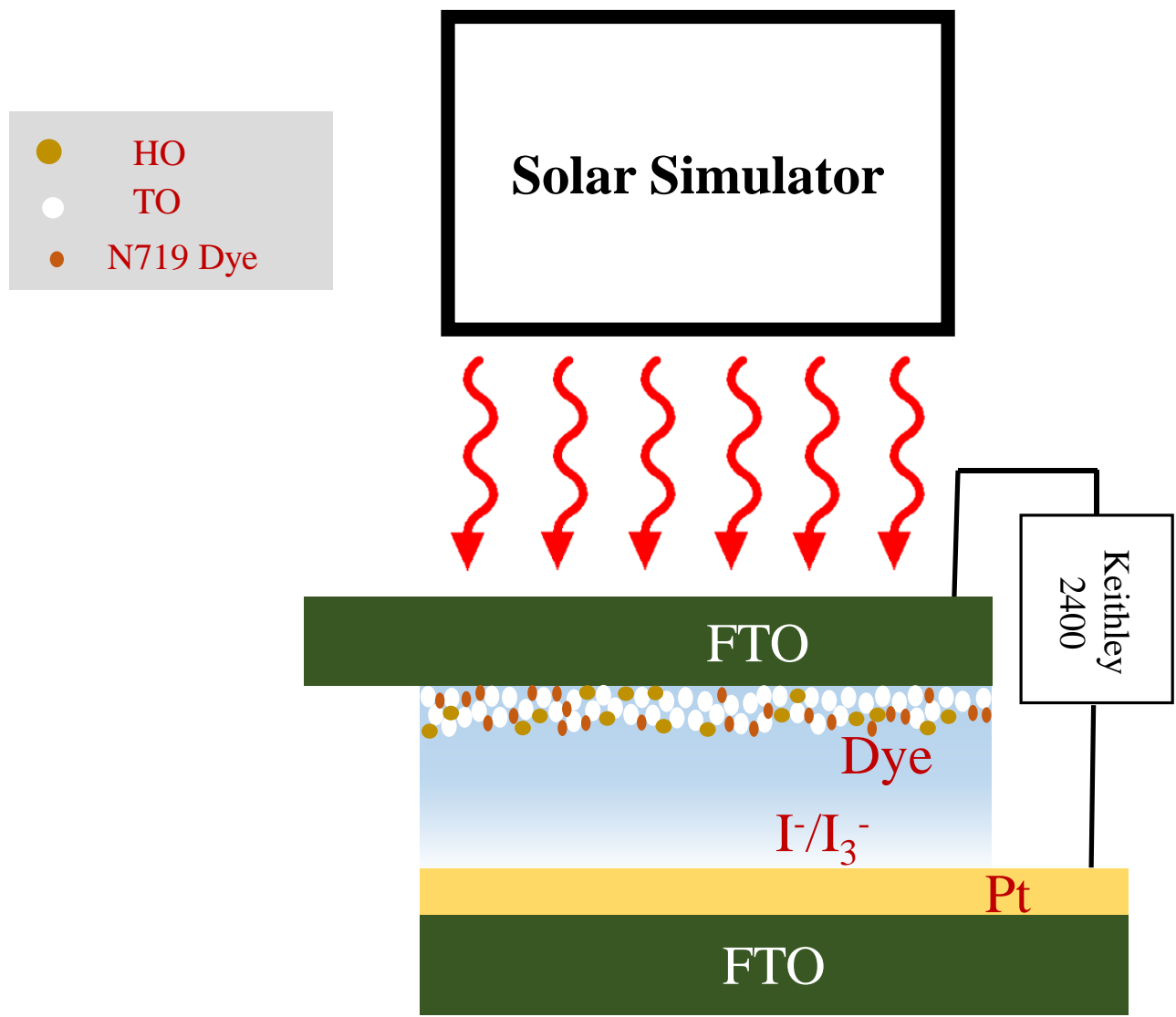

**Figure 2:** Cross sectional view of the device that we fabricated for J – V characteristics. Between two Fluorine doped tin oxide (FTO) plates, there exists a thin film which consists of TO (white circles), HO (gold circles) and N719 Dye (brown circles). In addition, the photosensitive layer is covered with the Iodide/Triiodide electrolyte ($I^-/I_3^-$) to regenerate the oxidized dye. Here we use Pt electrode as cathode.

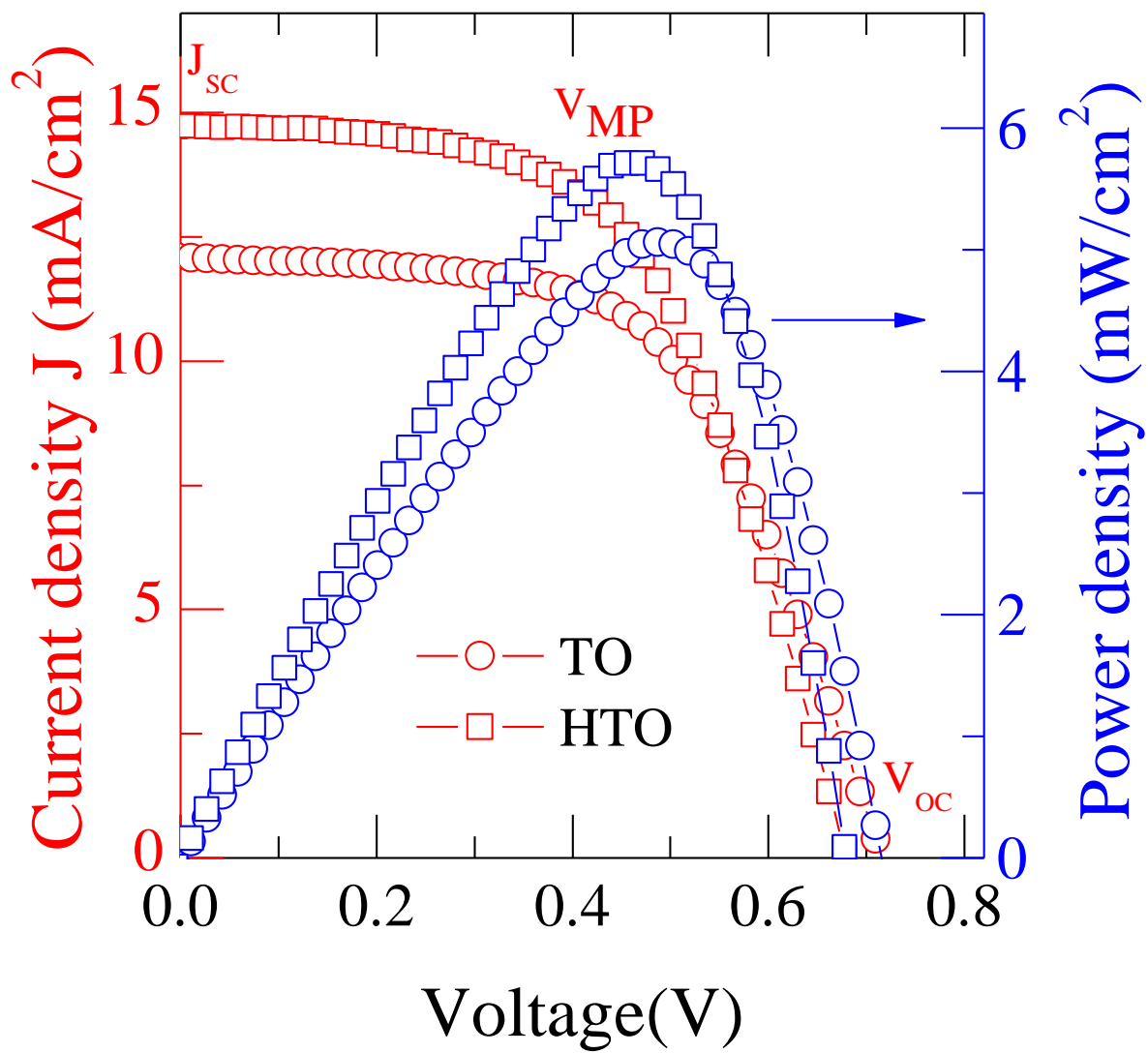

**Figure 3**: Voltage vs. Current density and Voltage vs. Power density of TO and HTO dye sensitized solar cells respectively. It is evident from the figure that current density as well as Power density increases with HO inclusion. In the graph, $J_{SC}$, $V_{MP}$ and $V_{OC}$ represent short circuit current density, voltage at maximum power point and open circuit voltage respectively.

**Table 1 :** Cell parameters of the fabricated devices and the corresponding methodology adopted for their determination are shown (See for instance Ref. no. 26)

| Parameter | Device | | Methodology used |
|---|---|---|---|
| | TO | HTO | (Ref. no. 26 ) |
| $V_{OC}$ (V) | 0.72 | 0.68 | Experiment |
| $J_{SC}$ (mA/cm$^2$) | 12.08 | 14.73 | |
| FF (%) | 58.59 | 57.21 | |
| η (%) | 5.22 | 5.90 | |
| Series resistance, $R_S$ (ohm cm$^2$) | 10.35 | 7.69 | $R_t = R_s + \dfrac{nk_BT}{q}\dfrac{1}{J_{sc}-J}$ |
| Shunt resistance, $R_{sh}$ (Kohm cm$^2$) | 2.27 | 3.68 | $J = \dfrac{(J_L+J_0)R_{sh}}{R_s+R_{sh}} + \dfrac{V}{R_s+R_{sh}}$ |
| Ideality factor, n | 2.28 | 2.67 | $R_r = \dfrac{nk_BT}{qJ_0}\exp(\dfrac{qV_P}{nk_BT})$ |
| Dark saturation current density, $J_0$ (mA/cm$^2$) | 5.68 x 10$^{-5}$ | 6.58 x 10$^{-4}$ | |

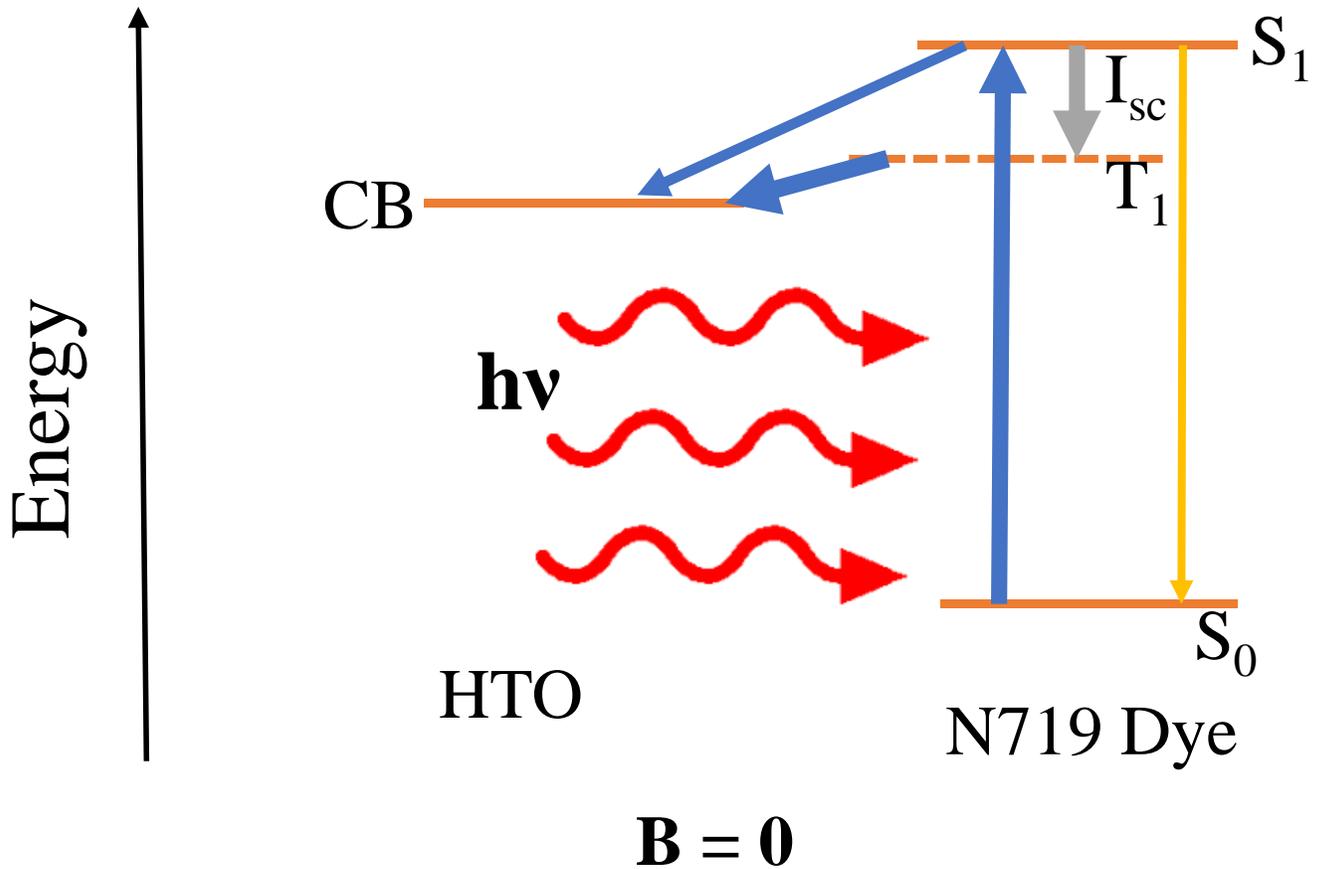

**Figure 4:** Photoexcitation of spin states in N719 dye without the presence of an external magnetic field. The corresponding charge transfer to the conduction band of TO dispersed with HO has also been shown. The blue and grey arrows indicate non- radiative transitions while the yellow arrows indicate radiative transitions. The thickness of arrows indicates the relative intensity of transitions. $S_0$, $S_1$ and $T_1$ represent the singlet ground state, singlet excited state and triplet excited states respectively. CB denotes the conduction band of TO and $I_{sc}$ indicates intersystem crossing.

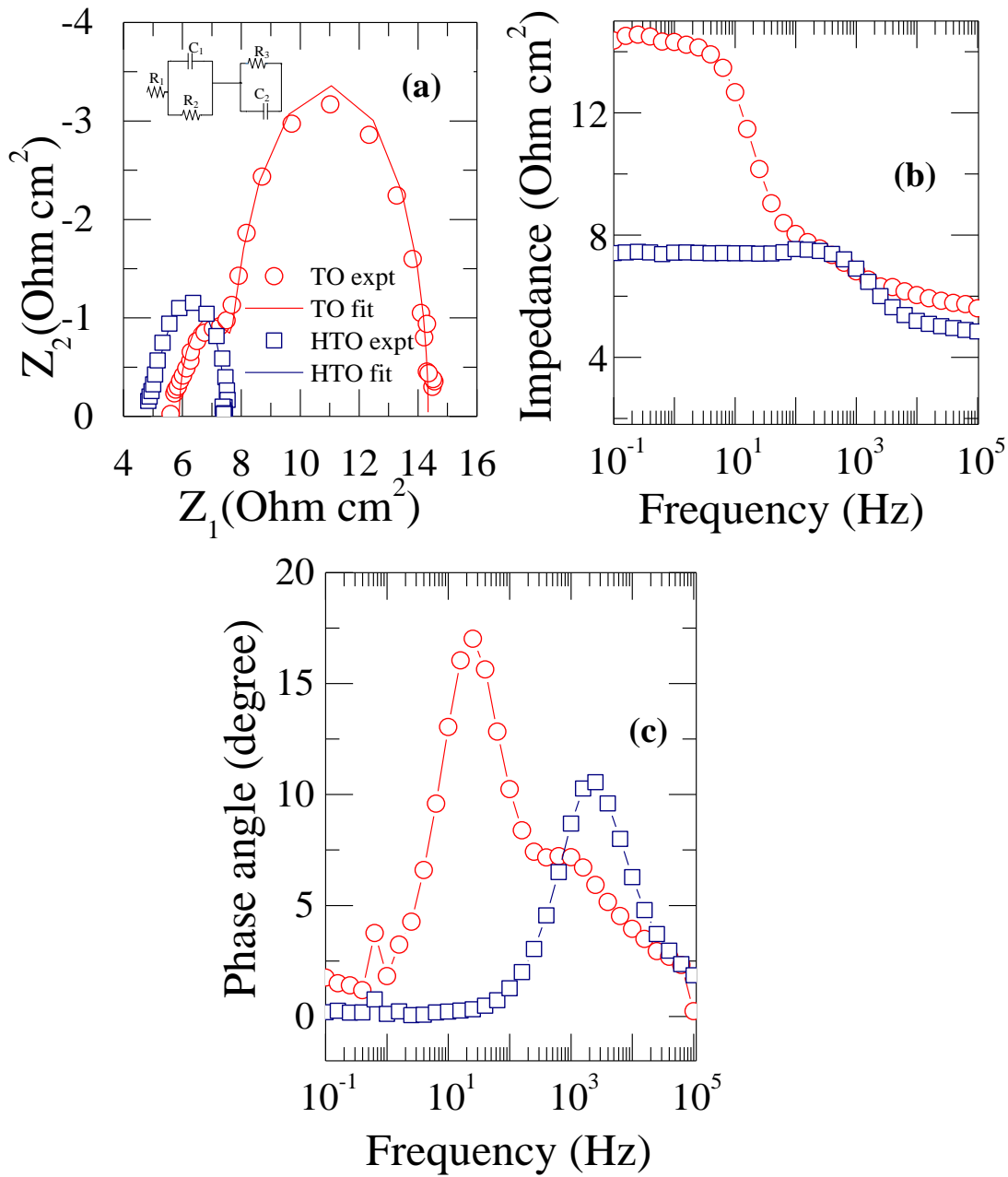

**Figure 5**: (a) Nyquist plot ((b) & (c)) Bode plots respectively for TO and HTO. Inset of (a) shows the electrical circuit used for EIS fitting. From Nyquist plot it is evident that radius of impedance arc decreases with HO due to the enhancement in the charge transport. From plot (b) it is clear that the impedance decreases with respect to the frequency, hinting that the real part dominates with respect to the frequency. Same is reflected as a decrease in phase angle from (c).

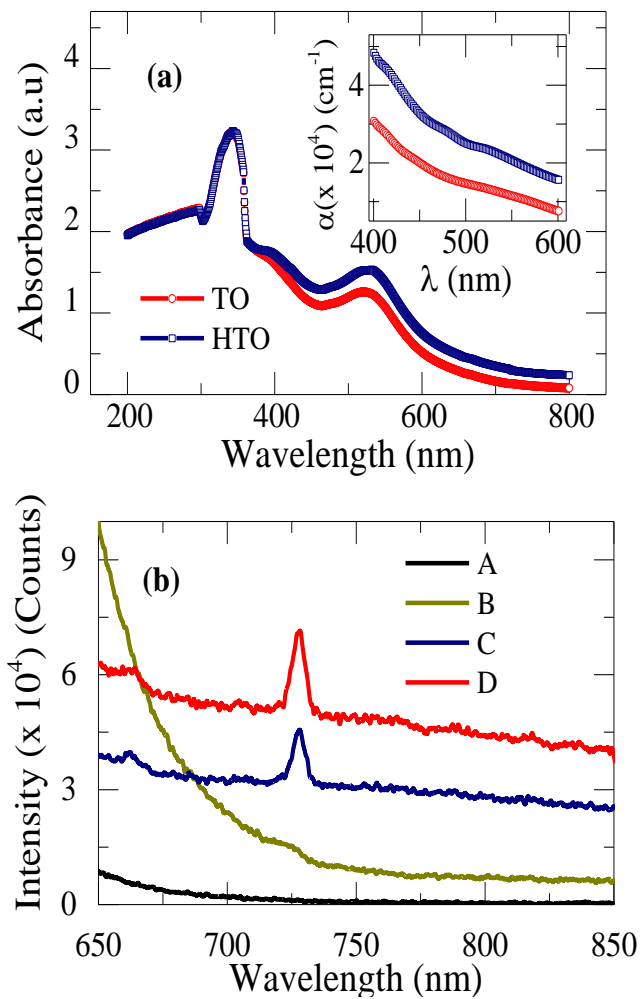

**Figure 6**: (a) Wavelength vs. absorbance spectra of TO and HTO. It is apparent that the absorption increases in the visible region upon inclusion of the HO. Inset shows the variation of absorption coefficient ($\alpha$) with wavelength. (b) Wavelength vs Photoluminescence intensity of a solution of methanol (A), a solution of 0.002 g $Ho_2O_3$ (HO) in methanol (B) and blade coated films of HTO (C) and TO (D) on FTO coated glass slides. A clear decrease in the emission intensity can be observed for sample C over D which corresponds to the increase in intersystem crossing with rare earth oxide addition

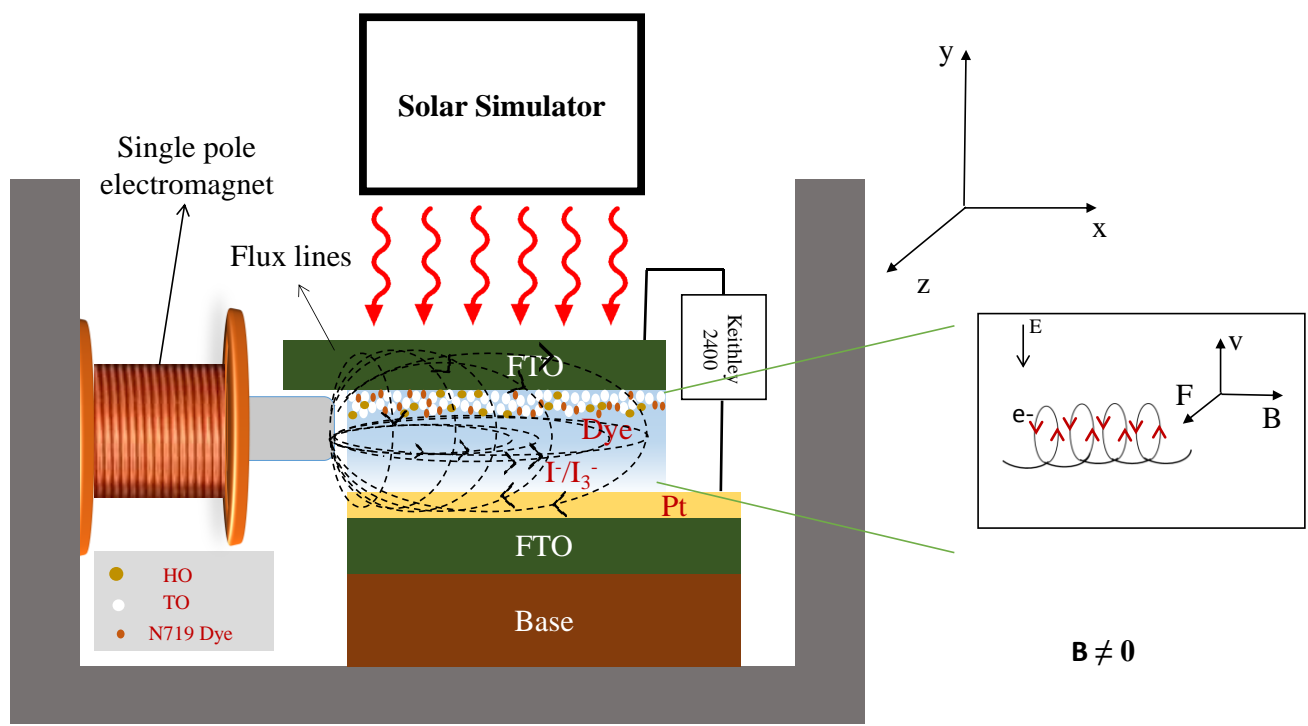

**Figure 7:** Schematic of the setup that we used to measure solar cell efficiency and other parameters. This schematic also demonstrates the kind of electromagnet that we used (single pole magnet) and the cross sectional view of the device that we fabricated. Between two Fluorine doped tin oxide (FTO) plates, there exists a thin film which consists of TO (white circles), HO (gold circles) and N719 Dye (brown circles). In addition, the photosensitive layer is covered with the Iodide/Triiodide electrolyte ($I^-/I_3^-$) to regenerate the oxidized dye. Here we use Pt electrode as cathode. The magnetic field has both the parallel and perpendicular components. Electric field is in (–) y – direction and magnetic field is in x – direction, both are perpendicular. As shown in inset, velocity of charge carriers is in y direction and magnetic field is in x – direction, the Lorentz force would be in z – direction. Electrons attain helical motion as shown in inset.

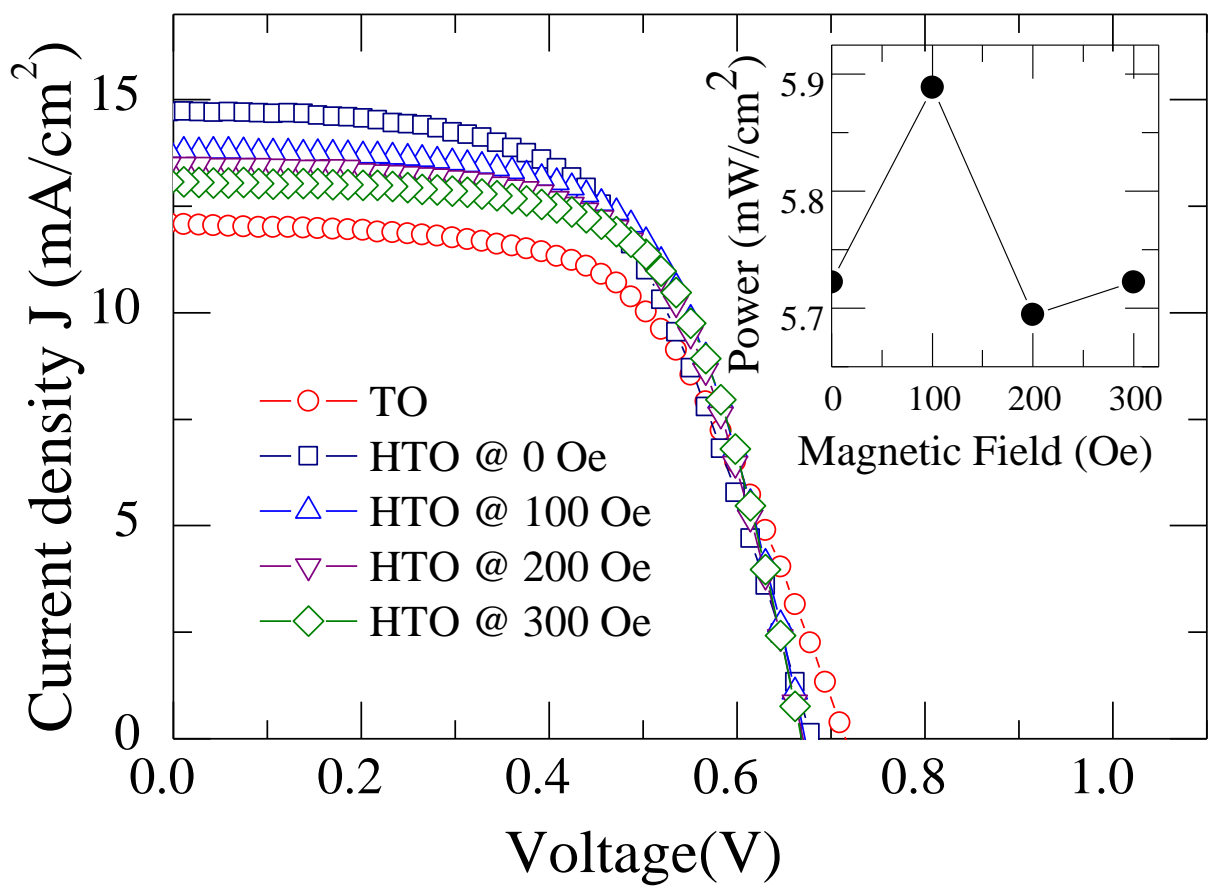

**Figure 8**: Current density (J) vs Voltage (V) curves for the fabricated DSSC devices with and without magnetic field. The applied magnetic field was varied from 0 Oersted (Oe) to 300 Oe in steps of 100 Oe. It is evident that power point enhanced for low magnetic fields.

**Table 2 :** Device parameters of TO , HTO devices and the corresponding methodology adopted for their determination are shown.

| Parameter | Device | | | | | Methodology used |
|---|---|---|---|---|---|---|
| | TO | HTO 0 Oe | HTO @100 Oe | HTO @200 Oe | HTO @300 Oe | (Ref. no. 26 ) |
| $V_{OC}$ (V) | 0.72 | 0.68 | 0.67 | 0.67 | 0.67 | Experiment |
| $J_{SC}$ (mA/cm$^2$) | 12.08 | 14.73 | 13.79 | 13.47 | 13.07 | |
| FF (%) | 58.59 | 57.21 | 63.43 | 63.19 | 65.55 | |
| η (%) | 5.22 | 5.90 | 6.06 | 5.88 | 5.90 | |
| Series resistance, $R_S$ (ohm cm$^2$) | 10.35 | 7.69 | 2.46 | 2.76 | 1.64 | $R_t = R_s + \dfrac{nk_BT}{q}\dfrac{1}{J_{sc} - J}$ |
| Shunt resistance, $R_{sh}$ (Kohm cm$^2$) | 2.27 | 3.68 | 4.82 | 2.93 | 3.48 | $J = \dfrac{(J_L + J_0)R_{sh}}{R_s + R_{sh}} + \dfrac{V}{R_s + R_{sh}}$ |
| Ideality factor, n | 2.28 | 2.67 | 2.97 | 2.88 | 3.54 | $R_r = \dfrac{nk_BT}{qJ_0}\exp(\dfrac{qV_P}{nk_BT})$ |
| Dark saturation current density, $J_0$ (mA/cm$^2$) | 5.68 x 10$^{-5}$ | 6.58 x 10$^{-4}$ | 1.83 x 10$^{-3}$ | 1.48 x 10$^{-3}$ | 4.89 x 10$^{-3}$ | |